\documentclass[%
 reprint,
 amsmath,amssymb,
 aps,
pra,
floatfix,
superscriptaddress]{revtex4-2}
\usepackage[T1]{fontenc} 
\usepackage[utf8]{inputenc}

\usepackage{graphicx}
\usepackage{dcolumn}
\usepackage{bm}

\usepackage{siunitx} 
\usepackage{hyperref}
\usepackage[all]{hypcap}
\hypersetup{colorlinks=true, allcolors=black}

\newcommand{\QMAffiliation}{%
Quantum Motion, 9 Sterling Way, London, N7 9HJ, United Kingdom
}

\begin{document}



\title{%
\texorpdfstring
{Fast On-Chip Thermometry with TiN Kinetic Inductance Resonators \\
in 22\hspace{0.1666em}nm Process Technology}
{Fast On-Chip Thermometry with TiN Kinetic Inductance Resonators in 22 nm Process Technology}
}

\author{Johann Drayne}
\affiliation{\QMAffiliation}

\author{George Ridgard}
\affiliation{\QMAffiliation}

\author{Lorenzo Peri}
\affiliation{\QMAffiliation}

\author{Frederic Schlattner}
\affiliation{\QMAffiliation}

\author{Fabio Olivieri}
\affiliation{\QMAffiliation}

\author{Mathieu de Kruijf}
\affiliation{\QMAffiliation}

\author{Isaac Harris}
\affiliation{\QMAffiliation}

\author{Grayson Noah}
\affiliation{\QMAffiliation}

\author{James Kirkman}
\affiliation{\QMAffiliation}

\author{Alberto Gomez-Saiz}
\affiliation{\QMAffiliation}
\affiliation{Department of Electrical and Electronic Engineering, Imperial College London, London SW7 2AZ, United Kingdom}

\author{M. Fernando Gonzalez-Zalba}
\email{fernando@quantummotion.tech}
\affiliation{\QMAffiliation}
\affiliation{CIC nanoGUNE Consolider, Tolosa Hiribidea 76, E-20018 Donostia-San Sebastian, Spain}
\affiliation{IKERBASQUE, Basque Foundation for Science, E-48011 Bilbao, Spain}

\author{Thomas Swift}
\email{tom@quantummotion.tech}
\affiliation{\QMAffiliation}

\begin{abstract}
Understanding the temporal and spatial dependence of temperature is critical in high-performance cryogenic devices. 
Typical thermometry techniques struggle to simultaneously combine high bandwidth, sensitivity and on-chip integration, limiting their ability to measure fast intra-device thermal fluctuations. 
Here, we demonstrate a time-resolved thermometry platform based on TiN kinetic inductance resonators integrated in a $22\,\mathrm{nm}$ FDSOI chip.  
By tracking temperature-dependent shifts in the resonant frequency, we achieve sub-millikelvin temperature sensitivity down to temperatures of $100\,\mathrm{mK}$. 
Time-resolved on-chip pulsed heating experiments as a function of distance reveal an onset delay, consistent with a quasi-ballistic heat propagation velocity of $3.9 \pm 0.1\,\mathrm{mm\,\mu s^{-1}}$. 
We also show that elevated temperatures increase net thermal conductance, shortening thermal relaxation times across all spatial separations. 
This behaviour manifests in two distinct regimes: a substrate-limited regime at $100\,\mathrm{mK}$, where cooling rates vary with heater distance, and a Kapitza boundary-limited regime at $400\,\mathrm{mK}$, where thermal relaxation becomes more spatially uniform.
These measurements demonstrate kinetic inductance thermometry’s ability to rapidly probe non-equilibrium temperature dynamics in cryogenic devices such as quantum processors.
\end{abstract}

\maketitle 

\section{Introduction}
Most leading quantum computing platforms including semiconductor spin~\cite{Veldhorst_2017,Petit_2020,Zalba_2021}, superconducting~\cite{Bardin_2019,Acharya_2022}, photonic~\cite{Dong_2021} and trapped ion~\cite{Brandl_2016,Pagano_2018} need to be operated at cryogenic temperatures and require complex high-frequency pulse sequences for control and readout. To facilitate continued increases in the number of interaction-capable qubits, cryo-electronic circuits represent an attractive option to address the wiring bottleneck, particularly when co-integrated directly on a qubit chip or via flip-chip bonding and interposers~\cite{Sebastiano_2025, Brennan_2025}. However, thermal management in these architectures is challenging. Integrated cryo-electronics can introduce significant power dissipation, while high-frequency readout and control pulses also contribute dynamic heating through the rapid charging and discharging of local gate electrodes~\cite{Undseth_2023}. At sub-Kelvin temperatures, reduced cooling power, lower thermal conductivities and suppressed heat capacities can lead to rapid local temperature spikes that degrade qubit performance. Understanding and mitigating these dynamic thermal effects requires integrated, high-frequency and sensitive thermometry capable of mapping time- and distance-dependent heating dynamics directly on-chip.


State-of-the-art local thermometry in quantum processors often employs metal–superconductor junctions, which are difficult to integrate in semiconductor platforms and are sensitive to magnetic fields~\cite{Gasparinetti_2015,EfeGumus_2022,Giazotto_2006}.
Previous work has demonstrated multiple integrated thermometry techniques using devices native to commercial-grade CMOS manufacturing such as diode, superconducting phase transition, gate resistance, Johnson noise thermometry and quantum dot thermometry~\cite{Noah2024a,DEKRUIJF2024,huizinga2022integrated,ridgard2025voltage}, but these are mostly limited to sub-MHz operation. Comparatively little literature exists on thermometry which is simultaneously CMOS-integrated, high-frequency and can be operated in the sub-Kelvin regime. Notable exceptions include Champain et al.~\cite{Champain_2024}, who demonstrate radio-frequency readout of a quantum dot in a semiconductor qubit architecture and Lvov et al.~\cite{Lvov_2025} who report thermometry measurements using a transmon qubit. 

In this paper, we present a high-frequency thermometry technique that exploits the temperature dependence of the kinetic inductance of TiN thin films~\cite{Day_2006,Doyle_2016} available in the 22FDX process~\cite{Swift_2025}. Embedded into an LC parallel circuit, these inductors enable sub-mK temperature resolution with sub-microsecond dynamic response. By applying controlled short heat pulses from a local transistor heater and fitting single-exponential cooling curves, we map the system's thermal response time $\tau$ across base fridge temperature ($T_{\mathrm{MXC}}$), heater power, maximum local temperature ($T_{\mathrm{max}}$) and heater-to-thermometer distance ($d$). These measurements isolate the dominant contributions of interface boundary resistance and spatial phonon scattering, providing insight into thermal dynamics at mK temperatures.

\section{kinetic inductance thermometers}

\begin{figure}[!htbp]
\includegraphics[width=\columnwidth]{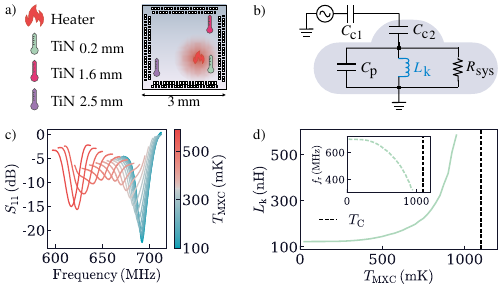}
\caption{\textbf{Device layout and temperature-dependent response.}
\label{fig:1_Measurement_Setup} 
(\textbf{a}) $22\,\mathrm{nm}$ FDSOI chip floorplan showing the approximate positions of three TiN thermometers and a transistor serving as a local heat source. Colour coding for the TiN ($0.2\,\mathrm{mm}$ green, $1.6\,\mathrm{mm}$ pink, $2.5\,\mathrm{mm}$ purple) is maintained throughout this work. 
(\textbf{b}) Equivalent resonant circuit diagram. The shaded region indicates on-chip components: the TiN kinetic inductor $L_{\mathrm{k}}$, system resistance $R_{\mathrm{sys}}$, parallel capacitance $C_{\mathrm{p}}$ (combining an on-chip resonator capacitor and intrinsic inductor parasitic capacitance) and line coupling capacitor $C_{\mathrm{c2}}$. Total line coupling is established via $C_{\mathrm{c2}}$ in series with an off-chip capacitor $C_{\mathrm{c1}}$. 
(\textbf{c}) Temperature dependence of the resonance in reflected microwave response magnitude ($S_{11}$) measured on the closest thermometer ($d = 0.2\,\mathrm{mm}$, green). The resonant frequency decreases with increasing temperature. 
(\textbf{d}) Kinetic inductance as a function of temperature for the $d = 0.2\,\mathrm{mm}$ thermometer. The dashed line indicates the critical temperature of the TiN thin film $T_{\mathrm{c}} = 1.1\,\mathrm{K}$. The inset shows the monotonic shift of the resonant frequency due to the temperature-dependent kinetic inductance.}
\end{figure}

The experiment is conducted on a $3\,\mathrm{mm} \times 3\,\mathrm{mm}$ chip fabricated using the GlobalFoundries 22FDX process~\cite{Carter_2016,Wang_2022}. A schematic of the chip is shown in Fig.~\ref{fig:1_Measurement_Setup}(a), with three TiN thermometers and a transistor acting as a heater. The external signals of all structures are directly connected to bond pads on the chip meaning no additional circuitry is needed to access them. 
The TiN film becomes superconducting below $T_{\mathrm{c}} \approx 1.1\,\mathrm{K}$ and exhibits a temperature-dependent kinetic inductance $L_{\mathrm{k}}(T)$. In the low-frequency limit ($\hbar \omega \ll 2\Delta$), $L_{\mathrm{k}}(T)$ is derived from the imaginary component of the Mattis–Bardeen complex conductivity~\cite{Mattis_1958, Gao_2008, Annunziata_2010}:

\begin{equation}
L_{\mathrm{k}}\left(T\right) = \frac{\hbar R_\square}{\pi \Delta\left(T\right) \tanh\left(\frac{\Delta\left(T\right)}{2 k_{\mathrm{B}} T}\right)} \frac{l}{w},
\label{eq:Lk_T_dependence}
\end{equation}

\noindent where $R_\square$ is the normal-state sheet resistance, $l$ is the strip length, $w$ is the strip width and $\Delta(T)$ is the temperature-dependent superconducting energy gap. 
Recalling that the gap closes near transition as $\Delta(T) \propto \sqrt{1 - T/T_{\mathrm{c}}}$, Eq.~\eqref{eq:Lk_T_dependence} captures the divergence of $L_{\mathrm{k}}(T)$ as $T \to T_{\mathrm{c}}$. 
While structural disorder in TiN films can introduce deviations from standard BCS theory~\cite{Driessen_2012}, this expression still provides an qualitative description of the temperature dependence.

The TiN film forms part of resonant $LC$ circuit Fig.~\ref{fig:1_Measurement_Setup}(b). The on-chip components (shaded region) consists of the kinetic inductance $L_{\mathrm{k}}$ (from the TiN film) in parallel with system resistance $R_{\mathrm{sys}}$ and parallel capacitance $C_{\mathrm{p}}$, which combines the on-chip resonator capacitor and intrinsic inductor parasitic capacitance. Line coupling is established by an on-chip capacitor $C_{\mathrm{c2}}$ in series with an off-chip capacitor $C_{\mathrm{c1}}$, yielding an effective coupling capacitance $C_{\mathrm{c}} = (C_{\mathrm{c1}}^{-1} + C_{\mathrm{c2}}^{-1})^{-1}$ and a total circuit capacitance $C_{\mathrm{total}} = C_{\mathrm{c}} + C_{\mathrm{p}} = 407\,\mathrm{fF}$ for the $d = 0.2\,\mathrm{mm}$ thermometer. As temperature rises, thermal Cooper-pair depairing increases $L_{\mathrm{k}}(T)$, shifting the resonant frequency lower Fig.~\ref{fig:1_Measurement_Setup}(c). 
Assuming kinetic inductance dominates over geometric inductance ($L \approx L_{\mathrm{k}}$), $L_{\mathrm{k}}$ is extracted directly from the resonant frequency $f_{\mathrm{r}}$:

\begin{equation}
L_{\mathrm{k}} = \frac{1}{(2\pi f_{\mathrm{r}})^2C_{\mathrm{total}}},
\label{eq:Lk_Fr}
\end{equation}

\noindent yielding $L_\mathrm{k}(0) = 131\,\mathrm{nH}$ for the thermometer located at $d = 0.2\,\mathrm{mm}$. Across the three TiN thermometers, intentional variations in coupling capacitance $C_{\mathrm{c}}$ alongside differences in baseline kinetic inductance ($131\,\mathrm{nH}$ and $149\,\mathrm{nH}$ across polyresistor variants~\cite{Swift_2025}) yield distinct base resonant frequencies, yet all thermometers maintain identical qualitative thermal behaviour. As shown in Fig.~\ref{fig:1_Measurement_Setup}(d), both $L_{\mathrm{k}}(T)$ and $f_{\mathrm{r}}(T)$ exhibit an increasingly steep gradient with rising temperature, providing high sensitivity as $T$ approaches $T_{\mathrm{c}}$.

Although atomic two-level systems (TLSs) in interfacial oxides can induce small non-monotonic frequency shifts at very low temperatures ($T \ll T_{\mathrm{c}}$)~\cite{Gao_tls_2008, Nicaise_2022, de_Ory_2025}, our measured frequency response remains strictly monotonic over the investigated temperature range ($T \ge 100\,\mathrm{mK}$), confirming that Cooper-pair thermal depairing is the dominant physical mechanism.




\section{Thermometer Calibration \& Benchmarking}


To calibrate the TiN thermometer, we step the mixing chamber temperature in $25\,\mathrm{mK}$ increments, allowing $15\,\mathrm{min}$ for thermal equilibration at each stage before measuring the reflected microwave response ($S_{11}$) using a Vector Network Analyzer (VNA), Fig.~\ref{fig:1_Measurement_Setup}(c). The resonant frequency $f_{\mathrm{r}}$, extracted from the minimum of the magnitude response is converted to $L_\mathrm{k}$ using Eq.~\ref{eq:Lk_Fr} and plotted as a function of temperature Fig.~\ref{fig:1_Measurement_Setup}(d). The mixing chamber temperature $T_{\mathrm{MXC}}$ is measured using a pre-calibrated RuOx thermometer.
Fitting the data to Eq.~\ref{eq:Lk_T_dependence} yields a critical temperature of $T_{\mathrm{c}} = 1.1\,\mathrm{K}$.

To evaluate performance and enable direct comparison with state-of-the-art cryogenic thermometry techniques, we calculate the bandwidth-normalised sensitivity $\delta T$~\cite{Noah2024a}, defined as,

\begin{equation}
\delta T =
\frac{1}{\sqrt{\mathrm{BW}}}\frac{\sigma_{f_\mathrm{r}}}{\partial f_\mathrm{r}/\partial T},
\label{eq:sensitivity}
\end{equation}

\noindent where $\sigma_{f_\mathrm{r}}$ is the standard deviation of the resonant frequency, $\partial f_{\mathrm{r}}/\partial T$ is the temperature responsivity and $\mathrm{BW}$ the effective measurement bandwidth. $\delta T$ represents the minimum resolvable temperature change for a given measurement bandwidth.

\begin{figure}[!htbp]
\includegraphics[width=\columnwidth]{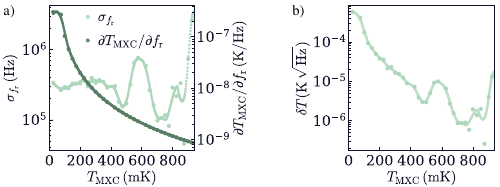}
\caption{\label{fig:2_SNBW}\textbf{Thermometer Bandwidth-Normalised Sensitivity (BWNS).}
(\textbf{a}) Resonant frequency uncertainty ($\sigma_{f_{\mathrm{r}}}$, light green) and temperature responsivity ($\partial f_{\mathrm{r}}/\partial T_{\mathrm{MXC}}$, dark green) as a function of mixing chamber temperature $T_{\mathrm{MXC}}$.
(\textbf{b}) Bandwidth-normalised sensitivity $\delta T$ versus $T_{\mathrm{MXC}}$. The order-of-magnitude enhancement in $\delta T$ at higher temperatures is driven primarily by the sharp increase in temperature responsivity rather than a reduction in frequency measurement uncertainty.}
\end{figure}

Benchmarking measurements are performed using a Qblox QRC module, which demodulates and digitises the in-phase ($I$) and quadrature ($Q$) components of the reflected signal. In the resonant frequency tracking mode, $I$ and $Q$ are recorded across a $7\,\mathrm{MHz}$ frequency span covering the resonator bandwidth, with a step size of $210\,\mathrm{kHz}$ (33 measurement points) to reconstruct $S_{11}(f)$. With an integration time of $1\,\mu\mathrm{s}$ per point, this yields an effective measurement bandwidth of $\mathrm{BW} \approx 30\,\mathrm{kHz}$.
Figure~\ref{fig:2_SNBW}(a) shows the resonant frequency uncertainty $\sigma_{f_{\mathrm{r}}}$, computed as the standard deviation across $250$ repeated measurements, alongside the responsivity $\partial f_{\mathrm{r}}/\partial T_{\mathrm{MXC}}$, which are then used to calculate $\delta T$ given in Fig.~\ref{fig:2_SNBW}(b). The significant enhancement in sensitivity at elevated temperatures is driven by the divergence of kinetic inductance near $T_{\mathrm{c}}$ (reflected in a sharply increasing $\partial f_{\mathrm{r}}/\partial T_{\mathrm{MXC}}$), whereas the measurement noise $\sigma_{f_{\mathrm{r}}}$ remains essentially flat.

Temperature variations modify both the reactive response of the superconductor (shifting $f_{\mathrm{r}}$ through Cooper pair depairing) and its dissipative response (altering the quality factor through quasiparticle generation (QP)), both of which are clearly visible in the shifting and broadening resonance curves in Fig.~\ref{fig:1_Measurement_Setup}(c). 
An alternative readout approach is to monitor the signal magnitude at a single fixed frequency, capturing both reactive and dissipative mechanisms simultaneously to provide enhanced sensitivity at low temperatures ($T < 100\,\mathrm{mK}$) where $\partial f_{\mathrm{r}}/\partial T$ saturates. 
Furthermore, because single-frequency acquisition requires only a single measurement point rather than the 33-point frequency sweep used to track $f_{\mathrm{r}}$, it could yield a factor of $\sqrt{33} \approx 6$ improvement in sensitivity ($\delta T$). 
However, resonant-frequency tracking offers a strictly monotonic temperature response, straightforward physical interpretation, and superior operational stability. We therefore employ resonant-frequency tracking for the remainder of this work.




\section{Time-Resolved Thermometry}

The high sensitivity of the TiN thermometer enables measurements of the dynamic temperature response of the chip to localised heat dissipation. To investigate this response, we employ a fast-pulse heater setup. 
The heater consists of a transistor with source-drain bias, $V_{\mathrm{SD}}$ applied through the source contact and the drain contact connected to ground. A bias tee on the transistor gate ($R=1\,\mathrm{M\Omega}$ and $C=0.57\,\mathrm{\mu F}$) enables simultaneous application of a DC offset voltage and high-frequency pulses. The heater power is calibrated by measuring the source current, $I_{\mathrm{S}}$ as a function of $V_{\mathrm{SD}}$ with the transistor gate fully turned on ($V_\mathrm{G}^\mathrm{DC} = 0.98\,\mathrm{V}$). 
During dynamic heating, $V_{\mathrm{SD}}$ is set to achieve the target power, while the gate is DC-biased just below threshold at $V_{\mathrm{G}}^{\mathrm{DC}} = 0.2\,\mathrm{V}$, ensuring near-zero source current prior to pulse arrival. Square voltage pulses of amplitude $V_{\mathrm{G}}^{\mathrm{Pulse}} = 0.78\,\mathrm{V}$ applied via a semi-rigid coaxial cable yield a net gate voltage of $V_{\mathrm{G}} = V_{\mathrm{G}}^{\mathrm{DC}} + V_{\mathrm{G}}^{\mathrm{Pulse}} = 0.98\,\mathrm{V}$ during pulse activation, driving the transistor into the fully turned-on state.

\begin{figure}[!htbp]
\includegraphics[width=\columnwidth]{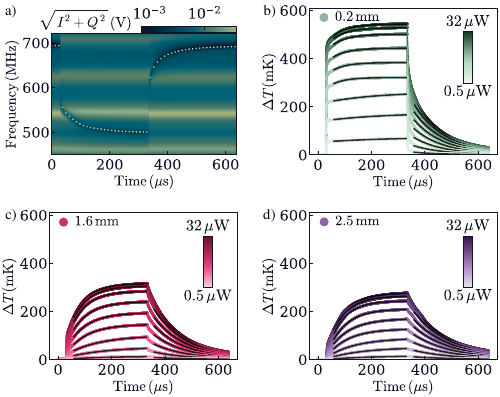}
\caption{\textbf{Time-resolved temperature response under pulsed heating.}
(\textbf{a}) Example measurement showing the time-resolved resonant frequency shift during a single stepped heating pulse ($P_{\mathrm{heat}} = 32\,\mu\mathrm{W}$). Markers highlight the extracted resonance as the TiN film warms and cools.
(\textbf{b}--\textbf{d}) Calculated temperature response across varying heater powers at heater distances of $0.2\,\mathrm{mm}$ (\textbf{b}), $1.6\,\mathrm{mm}$ (\textbf{c}) and $2.5\,\mathrm{mm}$ (\textbf{d}) from the heat source.}
\label{fig:3_ONOFF_Pulse} 
\end{figure}

An example measurement is shown in Fig.~\ref{fig:3_ONOFF_Pulse}(a) for a heating pulse with power $P_{\mathrm{heat}} = 32\,\mu\mathrm{W}$ and duration $t_{\mathrm{heat}} = 300\,\mu\mathrm{s}$. 
The map in this figure is constructed by recording the in-phase ($I$) and quadrature ($Q$) signals as a function of time, at a fixed RF frequency, repeated across a range of RF frequencies. 
Fifty traces are averaged together at each frequency. A $10\,\mathrm{ms}$ delay between heating pulses ensures full thermal relaxation back to baseline ($>60$ times the characteristic cooling time $\tau_{\mathrm{cooling}}$ discussed below). 
At each time step, the resonant frequency is extracted and converted to a temperature using a pre-calibrated frequency-temperature relation Fig.~\ref{fig:1_Measurement_Setup}(d). The resulting time-resolved resonant frequency during the heating pulse is indicated by the markers in Fig.~\ref{fig:3_ONOFF_Pulse}(a). The measurement is repeated for a range of heating powers and distances Fig.~\ref{fig:3_ONOFF_Pulse}(b)-(d). While reconstructing $f_{\mathrm{r}}(t)$ from the full frequency span ensures a monotonic mapping to temperature, analysing single-frequency $IQ$ traces directly offers potential sensitivity improvements toward the single-tone limit. Additionally, frequency-domain multiplexing could enable simultaneous acquisition across multiple RF tones, significantly reducing total measurement time.

To extract thermal timescales, the temperature response after the heater is turned off is fit using a single-exponential model,


\begin{equation}
T(t) = 
T_{\mathrm{base}} + \Delta T \, e^{-t/\tau_{\mathrm{cooling}}}, 
\label{eq:exponential_fits}
\end{equation}

\noindent where $T_{\mathrm{base}}$ is the baseline temperature and $\Delta T$ is the temperature rise above baseline. The fitting window starts $26\,\mu\mathrm{s}$ after pulse termination and spans a $280\,\mu\mathrm{s}$ duration. Traces recorded across heater powers $P$ are optimised simultaneously via a constrained global routine enforcing monotonic scaling of $\Delta T(P)$, with $T_{\mathrm{base}}$ anchored to the unheated pre-pulse baseline ($t \le 20\,\mu\mathrm{s}$).

To capture the primary thermal relaxation kinetics, we model the decay using a first-order lumped-element approximation with an effective time constant $\tau_{\mathrm{cooling}} = C_{\mathrm{total}} / G_{\mathrm{total}}$. Although temperature-dependent heat capacities $C(T)$ and thermal conductances $G(T)$ strictly introduce non-linearities along the cooling trajectory, parametrising the transient with a single time constant provides a robust measure of the average relaxation rate across parameter space without introducing unconstrained fitting parameters. Here, $C_{\mathrm{total}}$ and $G_{\mathrm{total}}$ represent the integrated lumped heat capacity of the chip assembly and the net thermal conductance to the cold bath, respectively.

These quantities can be expressed as the sum contribution from individual thermal elements.

\begin{equation}
\tau_{\mathrm{cooling}} = \frac{C_{\mathrm{Si,ph}} + C_{\mathrm{epoxy}} + C_{\mathrm{e}}}
{G_{\mathrm{Si,ph}} + G_{\mathrm{epoxy}} + G_{\mathrm{K_R}}} \approx \frac{c_3 T^3 + c_1 T}{g_3 T^3 + g_2 T^2},
\label{eq:tau_cooling}
\end{equation}

\begin{equation}
\frac{d\tau_{\mathrm{cooling}}}{dT} = \frac{c_3g_2T^2-c_1(2g_3T+g_2)}{T^2(g_3T+g_2)^2}.
\label{eq:tau_cooling_dt}
\end{equation}

Here, $c_n$ and $g_n$ are temperature-independent coefficients that parametrise the magnitude of the corresponding $T^n$ power-law contribution.
The total heat capacity $C_{\mathrm{total}}$ combines three distinct sources: (C.i)~3D Debye phonons in the crystalline silicon substrate ($C_{\mathrm{Si,ph}} \propto T^3$)~\cite{Ashcroft_1976}, (C.ii)~the silver epoxy layer ($C_{\mathrm{epoxy}}$) which incorporates both a cubic lattice phonon term ($T^3$) and a linear two-level system (TLS) term ($T$) stemming from structural disorder~\cite{Zeller_1971, Pohl_2002}, and (C.iii)~free electrons in metallic circuit components ($C_{\mathrm{e}} \propto T$)~\cite{Kittel_2005, Savin_2005}. 
These metallic components -- which include contiguous metal traces across the chip, interconnects and ground planes -- provide parallel electronic conduction pathways across the die, while acting as thermal reservoirs that absorb energy during a heating pulse and release it during cooling.
The total thermal conductance $G_{\mathrm{total}}$ is governed by three thermal resistances in series: (G.i)~bulk phonon transport through the silicon substrate ($G_{\mathrm{Si,ph}} \propto T^3$)~\cite{Pobell1996, Savin_2005}, (G.ii)~acoustic boundary mismatch between substrate-epoxy and epoxy-sink interfaces (Kapitza conductance, $G_{\mathrm{Kapitza}} \propto T^3$)~\cite{Little_1959, Lounasmaa_1974, Savin_2005} and (G.iii)~phonon scattering within the amorphous silver epoxy layer ($G_{\mathrm{epoxy}} \propto T^2$)~\cite{Zeller_1971, Pohl_2002, Savin_2005}. For thin epoxy layers ($d_{\mathrm{epoxy}} < 100\,\mu\mathrm{m}$) and sub-Kelvin temperatures, the substrate-sink resistance is typically dominated by the Kapitza resistance~\cite{Savin_2005}. The printed circuit board (PCB) carrying the chip is tightly bolted to the mixing chamber plate of the dilution refrigerator and is therefore assumed to be well thermalised to the base temperature. Analysing the temperature dependence of $\tau_{\mathrm{cooling}}$ across parameter space allows us to disentangle these competing contributions.

\begin{figure}[!htbp]
\includegraphics[width=\columnwidth]{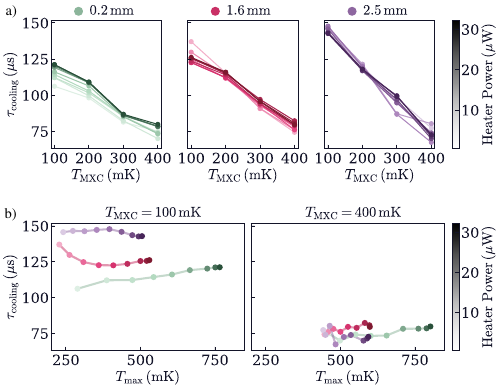}
\caption{\textbf{Thermal cooling times across parameter space.}
(\textbf{a}) Cooling time constant $\tau_{\mathrm{cooling}}$ versus mixing chamber temperature $T_{\mathrm{MXC}}$ for three heater distances ($d = 0.2\,\mathrm{mm}, 1.6\,\mathrm{mm}, 2.5\,\mathrm{mm}$) from the heat source across varying heater powers.
(\textbf{b}) $\tau_{\mathrm{cooling}}$ versus peak pulse temperature $T_{\mathrm{max}}$ at $T_{\mathrm{MXC}} = 100\,\mathrm{mK}$ (left) and $400\,\mathrm{mK}$ (right) across varying heater powers.}
\label{fig:cooling_tau}
\end{figure}

The extracted cooling time $\tau_{\mathrm{cooling}}$ reveals four distinct cooling behaviours Fig.~\ref{fig:cooling_tau}:

\paragraph{Thermal conductance limits the cooling time.}
Across all distances from the heater, $\tau_{\mathrm{cooling}}$ decreases monotonically from $120\mathrm{-}150\,\mu\mathrm{s}$ at $100\,\mathrm{mK}$ down to $75\,\mu\mathrm{s}$ at $400\,\mathrm{mK}$ Fig.~\ref{fig:cooling_tau}(a). 
The decrease in $\tau_{\mathrm{cooling}}$ with increasing $T_{\mathrm{MXC}}$ indicates that the temperature dependence of the cooling time is dominated by the increasing thermal conductance $G_{\mathrm{total}}$, rather than the corresponding increase in total heat capacity $C_{\mathrm{total}}$. 
To identify which terms give rise to the observed decrease in $\tau_{\mathrm{cooling}}$, we require $d\tau_{\mathrm{cooling}}/dT<0$. 
From Eq.~\eqref{eq:tau_cooling_dt}, this inequality is fulfilled when: (i)~linear heat capacities ($c_1 T$, originating from metallic interconnects and epoxy two-level systems) dominate over cubic lattice phonons ($c_3 T^3$), or (ii)~$T^3$ thermal conductance channels ($g_3 T^3$, incorporating bulk substrate transport $G_{\mathrm{Si,ph}}$ and interface Kapitza conductance $G_{\mathrm{Kapitza}}$) provide a sufficiently strong temperature dependence relative to the $T^2$ epoxy scattering term ($g_2 T^2$).


\paragraph{Distance-dependent cooling at $100\,\mathrm{mK}$.}
At this base temperatures, $\tau_{\mathrm{cooling}}$ increases with distance, from apx. $105\mathrm{-}120\,\mathrm{\mu s}$ at $0.2\,\mathrm{mm}$ to apx. $140\mathrm{-}150\,\mu\mathrm{s}$ at $2.5\,\mathrm{mm}$ as shown in Fig.~\ref{fig:cooling_tau}(a). 
Large temperature gradients persist across the chip as bulk substrate thermal resistance dominates over Kapitza boundary resistance, creating a lateral transport bottleneck that increases the cooling time at larger distances.

\paragraph{Distance-independent cooling at $400\,\mathrm{mK}$.}
In contrast, at an elevated temperatures, $\tau_{\mathrm{cooling}}$ is independent of distance ($d = 0.2\mathrm{-}2.5\,\mathrm{mm}$) and averages $75\,\mu\mathrm{s}$ across all applied powers Fig.~\ref{fig:cooling_tau}(a). 
Here, substrate thermal conductance increases sufficiently that global Kapitza boundary resistance takes over as the primary thermal bottleneck, causing the chip to relax as a single isothermal system.

\paragraph{Maximum temperature-dependent cooling.}
At $100\,\mathrm{mK}$, $\tau_{\mathrm{cooling}}$ increases with distance for a given maximum temperature $T_{\mathrm{max}}$ reached at pulse termination Fig.~\ref{fig:cooling_tau}(b).
Due to diffusive heat spreading, reaching an equivalent $T_{\mathrm{max}}$ at larger distances requires injecting significantly more total energy into the chip. 
Dissipating this additional stored energy through the substrate-limited transport channel takes longer, increasing the cooling time.
At $400\,\mathrm{mK}$, higher overall thermal conductance dissipates this heat rapidly, rendering $\tau_{\mathrm{cooling}}$ independent of $T_{\mathrm{max}}$.

Intrinsic superconducting relaxation processes could, in principle, contribute to the observed decay kinetics. Quasiparticle recombination exhibits an exponential temperature dependence, $\tau_{\mathrm{qp}} \propto \exp(\Delta / k_{\mathrm{B}} T)$~\cite{Kaplan_1976, devisser_2014, Gao_tls_2008, Nicaise_2022}. 
We do not observe a dominant exponential signature in $\tau_{\mathrm{cooling}}$ indicating either that the measured temperature window ($100\mathrm{-}400\,\mathrm{mK}$) is too narrow or intrinsic superconductor recovery is not the rate-limiting bottleneck. 
Furthermore, reported QP recombination times in TiN thin films are substantially faster ($< 100\,\mu\mathrm{s}$)~\cite{Coumou_2013}, than our measured recovery times ($\sim 75\mathrm{-}150\,\mu\mathrm{s}$). 
This further suggests that substrate thermal conductance and Kapitza interface resistance are the rate-limiting bottleneck for thermal relaxation.

\section{Phonon Propagation Dynamics}

The model used to fit data in Fig.~\ref{fig:3_ONOFF_Pulse}(b) qualitatively describes the long-time thermal response but does not capture the initial dynamics immediately following the start of the heating pulse. A rapid temperature rise occurs during the first $\mathrm{\sim1\,\mu s}$ before the system enters the slower equilibration regime described by Eq.~\ref{eq:exponential_fits}. To resolve this initial response, we perform measurements with improved temporal resolution using an integration time of $40\,\mathrm{ns}$.


\begin{figure}[!htb]
\includegraphics[width=\columnwidth]{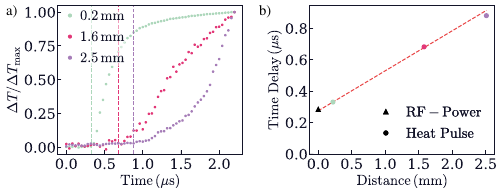}
\caption{\textbf{Time delay of system response to heating.}
(\textbf{a}) Time-resolved temperature response following a single stepped heating pulse step (applied at $\mathrm{t=0}$), measured at three distances from the heat source. Dashed lines mark the onset of the delayed temperature rise following the heating pulse, identified as the cross-over from an initial linear segment to exponential growth obtained from a bi-partite fit. Temperatures are normalised for clarity.
(\textbf{b}) Time delay versus distance from the heater. A straight line to the coloured markers yields a velocity $1/m=3.9\pm0.1\,\mathrm{mm\,\mu s^{-1}}$ and time offset $c=274\pm4\,\mathrm{ns}$. The black marker (not included in the fit) denotes the delayed response due to a change in RF power.}
\label{fig:4_40ns_delay} 
\end{figure}

Figure~\ref{fig:4_40ns_delay}(a) shows the normalised temperature response after a heating pulse begins. The onset of the delayed temperature rise marked by the dashed lines is identified using a bipartite fit consisting of an initial linear baseline followed by exponential growth. The extracted onset delay times scale linearly with thermometer distance from the heater Fig.~\ref{fig:4_40ns_delay}(b). A linear fit yields a propagation velocity of $3.9\pm0.1\,\mathrm{mm\,\mu s^{-1}}$ and a temporal delay of $t_0=274\pm4\,\mathrm{ns}$. 
The linear dependence indicates propagation of a heat pulse with approximately constant velocity, consistent with quasi-ballistic heat transport, where $v\approx d/t$~\cite{Guy_2025}. The extracted velocity ($3.9\,\mathrm{mm\,\mu s^{-1}}$) is lower than the fast longitudinal acoustic sound speed in bulk silicon ($8.4\mathrm{-}9.1\,\mathrm{mm\,\mu s^{-1}}$)~\cite{McSkimin_1953}. 
This reduction is expected where at sub-Kelvin temperatures thermal energy is predominantly carried by slower transverse acoustic phonon modes ($v_{\mathrm{TA}} = 4.7\mathrm{-}5.8\,\mathrm{mm\,\mu s^{-1}}$)~\cite{Kittel_2005}. Furthermore, specular and diffuse boundary reflections off the top and bottom surfaces of the chip increase the effective path length relative to the straight-line distance $d$, reducing the net propagation velocity.

The temporal intercept $t_0$ represents a delay prior to the distance-dependent phonon flight time. To disentangle systemic instrumentation latency from local device physics, we performed a direct RF loopback measurement, establishing a time of flight of $t \approx 180\,\mathrm{ns}$ along the measurement lines and room-temperature electronics. Adding the internal ring-up time of the resonator ($t_{\mathrm{r}} \approx 15\,\mathrm{ns}$) accounts for a combined RF total delay of $ 195\,\mathrm{ns}$. To independently verify the intrinsic (zero distance) response time, we conducted a control experiment in which the resonance frequency was perturbed directly by fast-gating the RF drive power. The resulting delay is shown as a black triangle in Fig.~\ref{fig:4_40ns_delay}(b). Note this data point is not included in the linear fit but is in agreement with the intercept, confirming that the residual offset is shared by both thermal and electronic excitation pathways. Subtracting the known RF line propagation and resonator ring-up latencies isolates a residual delay of $t_{\mathrm{qp}} \approx 80\,\mathrm{ns}$. We attribute this remaining delay to quasiparticle generation dynamics and pair-breaking conversion efficiency within the thin-film superconductor, which dictates the temporal onset of the kinetic inductance shift~\cite{Guy_2025}. \\

\section{Outlook}
In summary, we demonstrate a time-resolved on-chip thermometry platform using TiN kinetic-inductance resonators integrated with local transistor heaters fabricated in $22\,\mathrm{nm}$ FDSOI technology. By combining fast pulsed heating protocols with radiofrequency readout, we resolve sub-microsecond thermal dynamics, acoustic propagation delays and thermal relaxation kinetics at mK temperatures. These results offer a practical framework for mapping non-equilibrium heat transport in cryogenic integrated circuits. 
Future work will focus on three key directions. First, testing modified chip-to-package bonding interfaces will allow us to test whether lowering the boundary Kapitza resistance directly reduces relaxation timescales as predicted. 
Second, performing simultaneous thermometry during active qubit operation (using thermometers already co-integrated on the quantum chip) will enable real-time tracking of local temperature transients, directly quantifying thermal cross-coupling and correlating temperature spikes with gate errors~\cite{Undseth_2023}.
Finally, expanding to larger 2D arrays of multiplexed thermometers will map anisotropic phonon propagation across the die and help isolate individual heat capacity and conductance contributions. Together, these capabilities offer the data needed to guide thermally robust layouts for co-packaged cryo-CMOS and large-scale quantum processors.

\section{Acknowledgements}
We thank Joe Finney, Alexander Waterworth and Jonathan Warren of Quantum Motion for their technical support and thank Nigel Cave at Global Foundries for fruitful discussions.
Quantum Technologies [EP/S021582/1]. 
M.F.G.Z. acknowledges a UKRI Future Leaders Fellowship [MR/V023284/1].
A.G.-S. acknowledges an Industrial Fellowship from the Royal Commission for the Exhibition of 1851.

\section{Author Contributions}
J. D., I. H. and T. S. acquired the data. 
J. D., T. S., G. R., L. P. and M. d. K. analysed the data. 
T. S. and M. F. G. Z. conceived the experiment. 
T. S., G. R., F. S. and G. N. supervised the work. 
F. O. designed the integrated circuits. 
J. K. designed and validated the wiring infrastructure. 
J. K. validated the integrated circuits.
A. G. S. conceived and designed the superconducting structures.
All authors contributed to the writing of this manuscript.

\section{Competing interests}
The authors declare no competing interests. 

\appendix


%

\end{document}